\documentclass[aps,twocolumn,prd,floatfix,tightenlines,final,preprintnumbers,
               superscriptaddress,showpacs,showkeys]{revtex4}
\newcommand{\be}{\begin{equation}}
\newcommand{\ee}{\end{equation}}
\usepackage{color}
\usepackage{epsfig}
\usepackage{placeins}
\usepackage{axodraw}
\usepackage{pstricks}
\usepackage{graphicx}
\usepackage{amsmath}
\usepackage{theorem}
\usepackage{amssymb}
\usepackage{latexsym}
\usepackage{epic}
\usepackage{graphics}
\usepackage{rotating}
\begin{document}

\newcommand{\vp}{\vec{p}\hspace{0.5mm}'} 
\newcommand{\vsq}{\vec{p}^{\hspace{0.5mm}2}} 

\preprint{
\vbox{
\hbox{ADP-07-02/T642}
\hbox{Edinburgh 2007/5}
}}

\title[Meson form factors from lattice QCD]{Pseudoscalar and vector
  meson form factors from lattice QCD}

\author{J.N.~Hedditch}\affiliation{Department of Physics and
  Mathematical Physics and \\
  Special Research Centre for the Subatomic Structure of Matter, \\
  University of Adelaide, 5005, Australia}
\author{W.~Kamleh}\affiliation{Department of Physics and
  Mathematical Physics and \\
  Special Research Centre for the Subatomic Structure of Matter, \\
  University of Adelaide, 5005, Australia}
\author{B.G.~Lasscock}\affiliation{Department of Physics and
  Mathematical Physics and \\
  Special Research Centre for the Subatomic Structure of Matter, \\
  University of Adelaide, 5005, Australia}
\author{D.B.~Leinweber}\affiliation{Department of Physics and
  Mathematical Physics and \\
  Special Research Centre for the Subatomic Structure of Matter, \\
  University of Adelaide, 5005, Australia}
\author{A.G.~Williams}\affiliation{Department of Physics and
  Mathematical Physics and \\
  Special Research Centre for the Subatomic Structure of Matter, \\
  University of Adelaide, 5005, Australia}

\author{J.M.~Zanotti}\affiliation{School of Physics, University of
  Edinburgh, Edinburgh EH9 3JZ, UK}

\begin{abstract}
  We present a study of the pseudoscalar and vector meson
  form factors, calculated using the Fat-Link Irrelevant Clover (FLIC)
  action in the framework of Quenched Lattice QCD. Of particular
  interest is the determination of a negative quadrupole moment,
  indicating that the $\rho$ meson is not spherically symmetric.
\end{abstract}



\maketitle

\section{Introduction}

The important role that electromagnetic form factors play in our
understanding of hadronic structure has been well documented for more
than fifty years.
The reason for their popularity is that they encode information about
the shape of hadrons, and provide valuable insights into their
internal structure in terms of quark and gluon degrees of freedom.

Most of the attention, both experimentally and theoretically, has
focused on the electromagnetic form factors of the nucleon (see
Refs.~\cite{Gao:2003ag,Hyde-Wright:2004gh,Arrington:2006zm,Perdrisat:2006hj,deJager:2006nt}
for recent reviews).
The electromagnetic form factors of pseudoscalar mesons, especially
the pion, being the lightest QCD bound state, have also been studied
extensively
\cite{Woloshyn:1985in,Wilcox:1985bv,Woloshyn:1985vd,Wilcox:1985iy,Draper:1988bp,Draper:1988xv} in 
lattice QCD.
More recently, there is a renewed interest in calculating the pion form
factor on the lattice
\cite{Nemoto:2003ng,vanderHeide:2003kh,Abdel-Rehim:2004gx,Bonnet:2004fr,Capitani:2005ce,Hashimoto:2005am,Brommel:2006ww}.
This is especially timely considering the new \cite{Horn:2006tm} and
reanalysis of old \cite{Tadevosyan:2006yd} experimental data from
JLab.

The vector meson form factors, on the other hand, have received less
attention (see
Refs.~\cite{Samsonov:2003hs,Choi:2004ww,Braguta:2004kx,Aliev:2004uj,Bhagwat:2006pu}
for recent work).
Of particular interest is the quadrupole moment of the $\rho$ meson,
where theoretical determinations can disagree by as much as a factor
of two \cite{Bhagwat:2006pu}.
%
%
We aim to resolve this issue by performing the first direct
lattice calculation of the $\rho$-meson quadrupole form factor.
Charge and magnetic form factors are also calculated and from these we
extract the relevant static quantities, namely the mean square
charge-radius and magnetic moment.
We also analyse the dependence of light-quark contributions to these
form factors on their environment and contrast these with a new
calculation of the corresponding pseudoscalar-sector result.

Our aim is to reveal the electromagnetic structure of vector mesons
and to study to what extent the qualitative quark model picture is
consistent with quenched lattice QCD.
Interestingly, it has been shown in a lattice calculation by
Alexandrou et~al.~\cite{Alexandrou:2002nn} that the distribution of
charge in the vector meson is oblate, and therefore not consistent
with the picture of a quark anti-quark in relative S-wave.
By calculating the vector meson quadrupole form factor we make a
direct comparison with the findings of Ref.~\cite{Alexandrou:2002nn}.

For each observable we calculate the quark sector contributions
separately. Using this additional information we examine the
environmental sensitivity of the light-quark contributions to the
pseudoscalar and vector meson charge radii. We also evaluate the
dominance of the light quark contributions to the $K$ and $K^*$.

This paper builds on the preliminary work presented in Ref.~\cite{Lasscock:2006nh}.
In Section~\ref{sec:Mff} we introduce the theoretical formalism of
meson form factors, including the techniques required to extract them
from a lattice calculation.
Section~3 contains details of our lattice simulation, while in
Section~4 we present and discuss our results for both pseudoscalar and
vector mesons.
Finally, in Section~5 we summarise our findings and discuss future
work. 

\section{Theoretical formalism}

\subsection{Meson form factors}
\label{sec:Mff}

Meson form factors are extracted from matrix elements involving the
vector (electromagnetic) current
\begin{eqnarray}
\langle M(\vp)| J^\alpha |M(\vec{p})\rangle\,,
\label{eqn:ME}
\end{eqnarray}
where $M(\vec{p})$ ($M(\vp)$) denotes a meson state with initial (final)
momentum $\vec{p}$ ($\vp$). The momentum transfer is $q_\mu=(p_\mu^\prime
-p_\mu)$.

For a pion, the matrix element in Eq.~(\ref{eqn:ME}) is described by a
single form factor
\begin{eqnarray}
\langle \pi(\vp)| J^\alpha |\pi(\vec{p})\rangle = \frac{1}{2\sqrt{E_{\pi}(\vec{p})
    E_{\pi}(\vp)}} \, [p^\alpha + p^{\alpha \prime}] F_\pi(Q^2)\,,
\label{eqn:pion-ME}
\end{eqnarray}
where $Q^2=-q^2$ is the invariant momentum transfer, and the energy of the 
pion with momentum $\vec{p}$ is $E_{\pi}(\vec{p}) = \sqrt{m_{\pi}^{2} + \vsq}$.
The formula for the kaons are exactly analogous. 
The $\rho$-meson, on the other hand, is spin-1 and is described by three
form factors \cite{Brodsky:1992px},
\begin{eqnarray}
\label{eq:vector-current}
\left \langle \rho(\vp, s') | J^{\alpha} | \rho(\vec{p}, s)  \right \rangle
 &=&  \cr
 & & \hspace{-2.7cm} \frac{1}{2\sqrt{E_{\rho}(\vec{p})E_{\rho}(\vp)}} \epsilon^{\prime \star }_\tau(p',s')\
                                                                    J^{\tau \alpha \sigma}(p',p)\, \epsilon_\sigma(p,s)
\label{ad}
\end{eqnarray}
where $\epsilon$ and $\epsilon^\prime$ are
the initial and final polarisation vectors, respectively, and
\begin{widetext}
\begin{equation}
\label{Jexpression}
  J^{\tau \alpha \sigma}(p^\prime,p) =
 - \left\{ G_{1}(Q^2)\, g^{\tau \sigma} \, [p^\alpha + p^{\alpha\prime}]  
     +  G_2(Q^2)[g^{\alpha \sigma} q^\tau - g^{\alpha \tau}  q^\sigma ] 
    -  G_3(Q^2) \, q^\sigma q^\tau \  \frac{p^\alpha + p^{\alpha\prime
        }}{2m_{\rho}^2} \right\}\ .
\end{equation}
\end{widetext}
The covariant vertex functions $G_{1,2,3}$ can be rewritten in terms
of the Sachs charge, magnetic and quadrupole form factors \cite{Brodsky:1992px,Arnold:1980zj},
\begin{eqnarray}
  G_Q(Q^2) &=& G_1(Q^2) - G_2(Q^2) + (1 +\eta) G_3(Q^2) \\
  G_M(Q^2) &=& G_2(Q^2) \\
  G_C(Q^2) &=& G_1(Q^2) + \frac{2}{3} \eta\, G_Q(Q^2) \, ,
\end{eqnarray}
where $m_{\rho}$ is the mass of the vector-meson system calculated on
the lattice and $\eta = Q^2/ 4 m_{\rho}^2$.

The charge $q_{\rho}$, magnetic moment $\mu_{\rho}$, and quadrupole moment $Q_{\rho}$
are then extracted from $G_C$, $G_M$, and $G_Q$, respectively at zero
momentum transfer
\begin{eqnarray}
  e \, G_C(0) &=& q_{\rho} \label{eqn:charge}\\
  e \, G_M(0) &=& 2m_{\rho}\mu_{\rho} \label{eqn:magmom}\\
  e \, G_Q(0) &=& m^2_{\rho}Q_{\rho} \label{eqn:quadmom}\, .
\end{eqnarray}
The formulae for the $K^{*}$ are exactly analogous.


\subsection{Meson form factors on the lattice}

The matrix elements in Eqs.~(\ref{eqn:pion-ME}) and (\ref{ad}) are obtained from
ratios of three-point and two-point correlation functions
\begin{equation}
R^{\alpha}(p',p) = \sqrt{\frac{
\left<G^{\alpha}(\vp,\vec{p},t_2,t_1)\right> \,
\left<G^{\alpha}(\vec{p},\vp,t_2,t_1)\right>}{ 
\Large{<}G(\vp,t_2)\Large{>} \,
\Large{<}G(\vec{p},t_2)\Large{>}}}
\label{eqn:3pt-pi}
\end{equation}
for pseudoscalar mesons, and
\begin{equation}
R^{\alpha}_{\mu\nu}(p',p) = \sqrt{\frac{
\left<G^{\alpha}_{\mu\nu}(\vp,\vec{p},t_2,t_1)\right> \,
\left<G^{\alpha}_{\nu\mu}(\vec{p},\vp,t_2,t_1)\right>}{ 
\Large{<}G_{\mu\mu}(t_2,\vp)\Large{>} \,
\Large{<}G_{\nu\nu}(t_2,\vec{p})\Large{>}}}
\label{eqn:3pt-rho}
\end{equation}
for vector mesons. Note repeated indices are not summed over.

The two-point correlation function for the pseudoscalar mesons is
\begin{eqnarray}
{G}(t_2,\vec{p}) &=&  \sum_{\vec{x_2}}e^{-i\vec{p}\cdot\vec{x}_{2}}\
 \langle\Omega| \chi(x_{2})\chi^{\dagger}(0) |\Omega\rangle\ . 
\label{eqn:2pt-ps}
\end{eqnarray}
Similarly for the vector mesons,
\begin{eqnarray}
{G}_{\mu\nu}(t_2,\vec{p}) &=& \sum_{\vec{x_2}}e^{-i\vec{p}\cdot\vec{x}_{2}}\
\langle\Omega|    \chi_{\mu}(x_{2})   \chi_{\nu}^{\dagger}(0)         |\Omega\rangle\ .
\label{eqn:2pt-vec}
\end{eqnarray}
\begin{widetext}
The three-point correlation function for the pseudoscalar meson is
\begin{eqnarray}
G^{\alpha}(t_2,t_1,\vp,\vec{p})
&=& \sum_{\vec{x_1},\vec{x_2}} e^{-i \vp \cdot (\vec{x}_{2} - \vec{x}_{1})}e ^{-i \vec{p} \cdot \vec{x}_{1}}\
\langle \Omega |\
\chi(x_{2}) J^{\alpha}(x_{1}) \chi^{\dagger}(0)\
| \Omega \rangle \ .
\label{eqn:3pt-ps}
\end{eqnarray}
Similarly the three-point function for the vector meson is
\begin{eqnarray}
G^{\alpha}_{\mu\nu}(t_2,t_1,\vp,\vec{p})
&=& \sum_{\vec{x_1},\vec{x_2}} e^{-i \vp \cdot (\vec{x}_{2} - \vec{x}_{1})}e ^{-i \vec{p} \cdot \vec{x}_{1}}\
\langle \Omega |\
\chi_{\mu}(x_{2}) J^{\alpha}(x_{1}) \chi_{\nu}^{\dagger}(0)\
| \Omega \rangle \ .
\label{eqn:3pt-vec}
\end{eqnarray}
\end{widetext}
The Lorentz indices $\mu$ and $\nu$ are only present for the vector mesons,
while $\alpha$ is the index of the electromagnetic current.

The ratios in Eqs.(\ref{eqn:3pt-pi}) and (\ref{eqn:3pt-rho}) are constructed in
such a way as to remove the time-dependence and constants of normalisation from
the correlation functions at large time separations, $t_{1}$ and $t_{2}-t_{1}$.

These ratios differ subtly from previous work~\cite{Leinweber:1990dv}, in that
we are explicitly enforcing the parity of the terms through
the choice of momenta $(p',p)$ and $(p,p')$ vs $(p',p)$ and $(-p,-p')$. 
This requires two three-point propagators (with momentum-transfer $q$ and $-q$) for
each configuration. However with the well established technique of averaging over
$U$ and $U^{*}$ configurations \cite{Leinweber:1990dv,Boinepalli:2006xd}, there is no additional cost.
%


\subsubsection{$\pi$-meson case}

Since the pion has zero spin, the vertex is extraordinarily simple and takes 
the form given in Eq.~(\ref{eqn:pion-ME}). Here we show how this function is 
extracted from the ratio Eq.~\eqref{eqn:3pt-pi} by evaluating the correlation 
functions at large Euclidean times.

First we define the matrix elements as,
\begin{eqnarray}
\label{eq:matrix_elements-ps}
\langle    \Omega    |    \chi_{\mu}(0)   | \pi(\vec{p}) \rangle &=& {1 \over \sqrt{2E_{\pi}(\vec{p})}} \lambda_{\pi}(\vec{p}) \ ,\cr
\langle \pi(\vec{p}) | \chi_{\mu}^{\dagger}(0) |\Omega\rangle  &=& {1 \over \sqrt{2E_{\pi}(\vec{p})}} \bar{\lambda}_{\pi}(\vec{p})\ .
\end{eqnarray}
Here $\lambda_{\pi}(\vec{p})$ and $\bar{\lambda}_{\pi}(\vec{p})$ are the couplings of the 
interpolator to the pion with momentum $\vec{p}$ at the sink and source respectively.
The momentum dependence allows for the use of smeared fermion sources and sinks. The bar
allows for different amounts of smearing at the source and sink.
\begin{widetext}
By inserting a complete set of energy and momentum states into
Eq.~\eqref{eqn:2pt-ps}, we can show that at large Euclidean time,
\begin{eqnarray}
\label{eq:2pt_larget-ps}
\lim_{t_{2}\rightarrow \infty}{G}(t_2,\vec{p}) = {e^{-E_{\pi}(\vec{p})t_{2}} \over 2 E_{\pi}(\vec{p})}\lambda_{\pi}(\vec{p}) \bar{\lambda}_{\pi}(\vec{p})\ .
\end{eqnarray}
Following the same treatment, one can show that the three-point function Eq.~\eqref{eqn:3pt-ps} 
at large Euclidean time is
\begin{eqnarray}
\label{eq:3pt_larget-ps}
\lim_{t_{1},t_{2}-t_{1}\rightarrow \infty} G^{\alpha}(t_2,t_1,\vp,\vec{p}) &=&\
{e^{-E_{\pi}(\vp)(t_{2} - t_{1})}e^{-E_{\pi}(\vec{p})t_{1}} \over  2\sqrt{E_{\pi}(\vec{p})E_{\pi}(\vp)}}\
\lambda_{\pi}(\vp) \langle \pi(\vp)       | J^{\alpha}(0)     | \pi(\vec{p}) \rangle\ \bar{\lambda}_{\pi}(\vec{p})\ .
\end{eqnarray}
\end{widetext}

Substituting these expressions into Eq.~\eqref{eqn:3pt-pi} and using Eq.~\eqref{eqn:pion-ME}, the ratio
$R^{\alpha}(p',p)$ simply reduces to
\begin{eqnarray}
R^{\alpha}(p',p) = \frac{1}{2\sqrt{E_{\pi}(\vec{p}) E_{\pi}(\vp)}}\, [p^{\alpha} +
{p}^{\alpha\prime}]F_\pi(Q^2)\,,
\end{eqnarray}
such that the large Euclidean time limits of the ratio $R_{\alpha}$ is a direct
measure of $F_\pi(Q^2)$ up to kinematical factors.


\subsubsection{$\rho$-meson case}

Following \cite{Brodsky:1992px}, we define the matrix element of the
electromagnetic current for $\rho$-meson in terms of the covariant vertex 
functions $G_{1,2,3}$ as in Eqs.~(\ref{ad}) and (\ref{Jexpression}).

The analogues of the matrix elements in Eq.~\eqref{eq:matrix_elements-ps} are
\begin{eqnarray}
\langle\Omega|\, \chi_\mu(0) \,| \rho(\vec{p}, s) \rangle  &=& {1 \over \sqrt{2E_{\rho}(\vec{p})}}\lambda_{\rho}(\vec{p})
\epsilon_{\mu} (p, s)  \cr
\langle \rho(\vec{p}, s) |\, {\chi_{\nu}}^{\dagger}(0)\, | \Omega\rangle  &=& {1 \over \sqrt{2E_{\rho}(\vec{p})}}\
\bar{\lambda}_{\rho} (\vec{p})\, \epsilon^{\star}_{\nu} (p, s) \ .
\end{eqnarray}
The polarisation vectors obey the transversality condition
\begin{eqnarray}
\label{eq:transversality}
\sum_s \epsilon_{\mu} (p, s)\, \epsilon^{\star}_{\nu} (p, s) = -
\left(g_{\mu\nu} - \frac{p_\mu p_\nu}{m_{\rho}^2} \right)\ ,
\end{eqnarray}
because the vector meson current is  a conserved current.

\begin{widetext}
The evaluation of the two- and three-point functions proceeds as for our
discussion of the pion. However the completeness relation includes a
sum over spin-states. Using the transversality condition
Eq.~\eqref{eq:transversality} the analogue of Eq.~\eqref{eq:2pt_larget-ps}
becomes
\begin{eqnarray}
\label{eq:2pt-larget-rho}
\lim_{t_{2}\rightarrow \infty}{G}(t_2,\vec{p}) &=& \sum_{s}\
                                             {e^{-E_{\rho}(\vec{p})t_{2}}\over 2E_{\rho}(\vec{p})}\
                                             \lambda_{\rho}(\vec{p}) \bar{\lambda}_{\rho}(\vec{p})
                                             \epsilon_{\mu} (p, s)\, \epsilon^{\star}_{\nu} (p, s)\cr
                                           &=& - {e^{-E_{\rho}(\vec{p})t_{2}} \over 2E_{\rho}(\vec{p}) }\
                                                 \lambda_{\rho}(\vec{p}) \bar{\lambda}_{\rho}(\vec{p})\
                                                 \left(g_{\mu\nu} - \frac{p_\mu p_\nu}{m_{\rho}^2} \right)\ .       
\end{eqnarray}
Similarly using Eq.~\eqref{eq:vector-current} we can evaluate the three-point function, 
\begin{eqnarray}
\label{threepointexpr}
\label{eq:3pt-larget-rho}
\lim_{t_{1},t_{2}-t_{1}\rightarrow \infty} G^{\alpha}_{\mu\nu}(t_2,t_1,\vp,\vec{p}) &=&\
\sum_{s,s'}{ e^{-E_{\rho}(\vp)(t_{2}-t_{1})}e^{- E_{\rho}(\vec{p}) t_{1}}\over 4E_\rho(\vec{p}) E_{\rho}(\vp) }\
\lambda_{\rho}(\vp) \epsilon_{\mu} (p', s')\
\epsilon^{\prime \star }_\tau(p',s')  J^{\tau \alpha \sigma}(p',p) \epsilon_\sigma(p,s) \
\bar{\lambda}_{\rho} (\vec{p})\, \epsilon^{\star}_{\nu} (p, s) \cr
&=&   { e^{-E_{\rho}(\vp)(t_{2}-t_{1})}e^{- E_{\rho}(\vec{p}) t_{1}}\over 4E_\rho(\vec{p}) E_{\rho}(\vp) }\
      \lambda_{\rho}(\vp) \bar{\lambda}_{\rho}(\vec{p})\
     \left( g_{\mu\tau} - \frac{p'_\mu p'_\tau}{m_{\rho}^2} \right)\
      J^{\tau \alpha \sigma}\left(  g_{\sigma \nu} - \frac{p_\sigma p_\nu}{m_{\rho}^2} \right) \ .
\end{eqnarray}
\end{widetext}

Inserting the above expressions into the ratio in
Eq.~(\ref{eqn:3pt-rho}), together with our choice of momentum used in
the simulations, namely $p' = (E_{\rho},p_x,0,0)$ $(E_{\rho} =
\sqrt{m_{\rho}^{2} + p_x^{2}})$ and $p = (m_{\rho},0,0,0)$, it is
possible to express $R^{\alpha}_{\mu\nu}$ in terms of the Sachs form
factors,
\begin{eqnarray*}
  {R}^{0}_{11} &=& \frac{p_x^2}{3 m_{\rho} \sqrt{E_{\rho} m_{\rho}} } G_Q(Q^2) + \frac{E_{\rho}+m_{\rho}}{2\sqrt{E_{\rho}m_{\rho}}}
G_C(Q^2)\,, \\
  {R}^{0}_{22} = {R}^{0}_{33} &=& - \frac{p_x^2}{6 m_{\rho} \sqrt{E_{\rho}m_{\rho}}} G_Q (Q^2) + 
\frac{E_{\rho}+m_{\rho}}{2 \sqrt{E_{\rho} m_{\rho}}} G_C(Q^2)\,, \\
  {R}^{3}_{13} = {R}^{3}_{31} &=& \frac{p_x}{2 \sqrt{E_{\rho}m_{\rho}}} G_M(Q^2) \, .
\end{eqnarray*}

The individual form factors are isolated as follows:
\begin{eqnarray}
G_C(Q^2) &=& \frac{2}{3} \frac{\sqrt{E_{\rho}m_{\rho}}}{E_{\rho}+m_{\rho}} \left( {R}^{0}_{11} + 
{R}^{0}_{22} + {R}^{0}_{33} \right)\,, \\
G_M(Q^2) &=& \frac{\sqrt{E_{\rho}m_{\rho}}}{p_x} \left( {R}^{3}_{13} + {R}^{3}_{31}
\right)\,, \\
G_Q(Q^2) &=& \frac{m_{\rho} \sqrt{E_{\rho}m_{\rho}}}{p^2_x} \left( 2 {R}^{0}_{11} - 
{R}^{0}_{22} - {R}^{0}_{33} \right)\,.
\label{eqn:sachsquadR}
\end{eqnarray}
While we have used the subscript $\rho$ to denote a vector meson, the 
results are applicable to vector mesons in general, including the 
$K^{*}$ for example.

\subsection{Extracting static quantities}

The mean squared charge radius $\langle r^2 \rangle$ is obtained from
the charge form-factor through the following relation,
\begin{equation}
\label{eqn:rsqeqn}
\langle r^2 \rangle = -6 \frac{\partial}{\partial Q^2} G(Q^2) {\Big
|}_{Q^2=0}\,.
\end{equation}
To calculate the derivative the monopole form is used,
\begin{equation}
\label{eqn:monopole}
G_C(Q^2) = \left ( \frac{1}{\frac{Q^2}{\Lambda^2}+1}\right)\ .
\end{equation}
$\Lambda$ is referred to as the monopole mass.
Inserting this form into Eq.~(\ref{eqn:rsqeqn}) and rearranging provides 
%
\begin{equation}
\langle r^2 \rangle = \frac{6}{Q^2}\left( \frac{1}{G_C(Q^2)} - 1  \right)\ ,
\end{equation}
valid for quantities with $G_C(Q^2=0) = 1$.

As mentioned in Sec.~\ref{sec:Mff}, the charge
(Eq.~(\ref{eqn:charge})), magnetic moment (Eq.~(\ref{eqn:magmom})),
and quadrupole moment (Eq.~(\ref{eqn:quadmom})) can be extracted from
the Sachs form factors at zero momentum transfer.  Since we perform
our calculations at a single, finite value of $Q^2$, we will need to
adjust our results to zero momentum transfer.

From studies of nucleon properties, it is observed that $G_M$ and $G_C$
have similar $Q^2$-scaling at small $Q^2$~\cite{Litt:1969my}.  In the
following, we shall assume that this scaling also holds for quark
contributions to mesons. If $G_C(0) = 1$,
we have
\begin{equation}
G_M(0) \simeq \frac{G_M(Q^2)}{G_C(Q^2)} \,.
\end{equation}

Whilst a similar scaling could be used to relate our quadrupole
form-factor to the quadrupole moment, we believe that the form-factor
at our small finite $Q^2$ ( $\simeq 0.22 {\rm GeV}$) will be of
greater phenomenological interest.  We note that for a positively
charged meson a negative value of $G_Q$ corresponds to an oblate
deformation.

\section{Method}

The electromagnetic form factors are obtained using the three-point
function techniques established by Leinweber, {\it et al.} in
Refs.~\cite{Leinweber:1990dv,Leinweber:1992hy,Leinweber:1992pv} and
updated for smeared sources in Ref.~\cite{Boinepalli:2006xd}.
Our quenched gauge fields are generated with the ${\mathcal O}(a^2)$
mean-field improved Luscher-Weisz plaquette plus rectangle gauge
action \cite{Luscher:1984xn} using the plaquette measure for the mean
link.  We use an ensemble of 379 quenched gauge field configurations on
$20^3 \times 40$ lattices with lattice spacing $a= 0.128$ fm. The
gauge field configurations are generated via the Cabibbo-Marinari
pseudo-heat-bath algorithm~\cite{Cabibbo:1982zn} using a parallel
algorithm with appropriate link partitioning \cite{Bonnet:2000db}.

We use the fat-link irrelevant clover (FLIC) Dirac operator
\cite{Zanotti:2001yb} which provides a new form of nonperturbative
${\mathcal O}(a)$ improvement \cite{Zanotti:2004dr}.  The improved
chiral properties of FLIC fermions allow efficient access to the light
quark-mass regime \cite{Boinepalli:2004fz}, making them ideal for
dynamical fermion simulations now underway \cite{Kamleh:2004xk}.
For the vector current, we an ${\cal O}(a)$-improved FLIC
conserved vector current \cite{Boinepalli:2006xd}. We use a 
smeared source at $t_{2} = 8$. Complete simulation details are described in 
Ref.~\cite{Boinepalli:2006xd}.

Table~\ref{table:parameters} provides the $\kappa$-values used in our
simulations, together with the calculated pseudoscalar and vector
meson masses.
While we refer to $m_\pi^2$ in our figures and tables to infer the
quark masses, we note that the critical value where the pion mass
vanishes is $\kappa_{\rm cr}=0.13135$. Importantly the vector mesons
remain bound at all quark masses considered in this calculation
due to finite volume effects. 
That is, the mass of the vector mesons is less than the energy
of the lowest lying multi-hadron state with the appropriate 
quantum numbers.

The strange quark mass is chosen to be the third heaviest quark mass.  
This provides a pseudoscalar mass of 697 MeV which compares well with the
experimental value of $( 2M_K^2 - M_\pi^2 )^{1/2} = 693\, {\rm MeV}$
motivated by chiral perturbation theory. 
Two vector-meson interpolating fields are considered, namely $\bar{q} \gamma_i q$ and $\bar{q}\gamma_i \gamma_4 q$. 
Since results for the two interpolators agree, we simply present the results for
the $\bar{q} \gamma_i q$ interpolator, which displays a significantly stronger
signal.

The error analysis of the correlation function ratios is performed via
a second-order, single-elimination jackknife, with the $\chi^2$ per
degree of freedom $(\chi^{2}_{\rm dof})$ obtained via covariance
matrix fits.
We perform a series of fits through the ratios after the current insertion at $t=14$. 
By examining the $\chi^2_{\rm dof}$ we are able to establish a valid
window through which we may fit in order to extract our observables.
In all cases, we required a value of $\chi^2_{\rm dof}$ no larger than 1.5.
The values of the static quantities quoted in this paper on a per
quark-sector basis correspond to values for single quarks
of unit charge.

\begin{table}[tbph]
\caption{Meson masses for the respective values of the 
hopping parameter $\kappa$.}
\label{table:parameters}
\begin{ruledtabular}
\begin{tabular}{lllll}
\noalign{\smallskip}
$\kappa$  & $am_\pi$  & $am_{K}$  & $am_{\rho}$  & $am_{K^*}$\\
\hline
\noalign{\smallskip}
$0.12780$  & $0.5411(10)$ & $0.4993(11)$  & $0.7312(30)$   & $0.7057(27)$  \\
$0.12830$  & $0.5013(11)$ & $0.4782(11)$  & $0.7067(36)$   & $0.6933(40)$  \\
$0.12885$  & $0.4539(11)$ & $0.4539(11)$  & $0.6797(46)$   & $0.6796(46)$  \\
$0.12940$  & $0.4014(12)$ & $0.4285(11)$  & $0.6537(49)$   & $0.6668(47)$  \\
$0.12990$  & $0.3471(15)$ & $0.4044(12)$  & $0.6309(56)$   & $0.6556(50)$  \\
$0.13205$  & $0.3020(19)$ & $0.3862(13)$  & $0.6160(64)$  & $0.6484(52)$  \\
$0.13060$  & $0.2412(42)$ & $0.3671(19)$  & $0.6039(71)$   & $0.6423(54)$  \\
$0.13080$  & $0.1968(52)$ & $0.3574(16)$  & $0.5982(80)$  & $0.6393(56)$  \\
\end{tabular}
\end{ruledtabular}
\end{table}

\section{Results}
\subsubsection{Charge radii}
\begin{figure}[b]
\begin{center}
\includegraphics[height=1.0\hsize,angle=90]{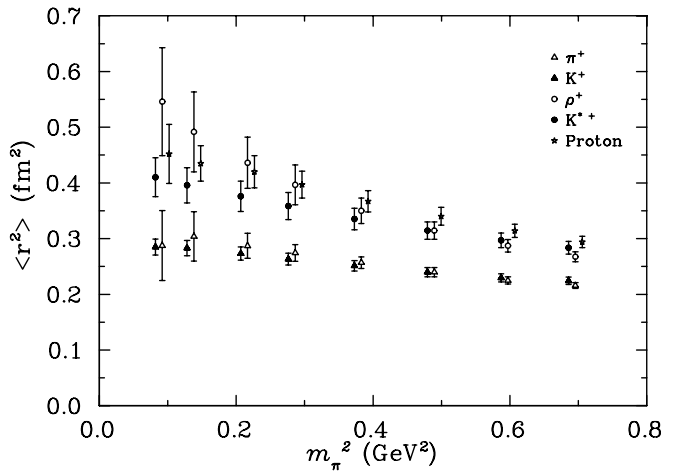}
\caption{\label{fig:rsquaredchargedsummary}
Strange and non-strange meson mean squared charge radii for charged
pseudoscalar and vector mesons. We also include for comparison results for the proton taken from Ref.~{\cite{Boinepalli:2006xd}}.
The $\pi$ and $\rho$-meson results are centred
on the relevant value of $m_\pi^2$, other symbols are offset horizontally for clarity.
}
\end{center}
\end{figure}

We begin the discussion of our results with the charge radii of the
vector and pseudoscalar mesons. From the quark model we would expect
a hyperfine interaction between the quark and anti-quark of the form
$\frac{\vec{\sigma_{q}}\cdot\vec{\sigma}_{\bar{q}}}{m_q m_{\bar{q}}}$.
The interaction is repulsive where the spins are aligned, as
in the vector mesons, and attractive where the spins are
anti-aligned, as in the pseudoscalar mesons.  
In Fig.~\ref{fig:rsquaredchargedsummary} we show the charge radii of the vector and pseudoscalar mesons.
For comparison the charge radius of the proton is also shown.
Indeed we find that the charge radii of the vector mesons are
consistently larger than the pseudoscalar mesons, and in fact similar
to the charge radii of the proton, even at heavier quark masses.
This is contrary to earlier lattice simulations with relatively small spatial extent~\cite{Draper:1989pi}, that have suggested that the $\pi^+$, $\rho^+$ and proton should have a very similar RMS charge radius at larger quark masses.
It is possible that the agreement obtained in the previous study reflects finite-volume effects attendant with the use of a small spatial volume. 

\begin{figure}[ht]
\includegraphics[height=1.0\hsize,angle=90]{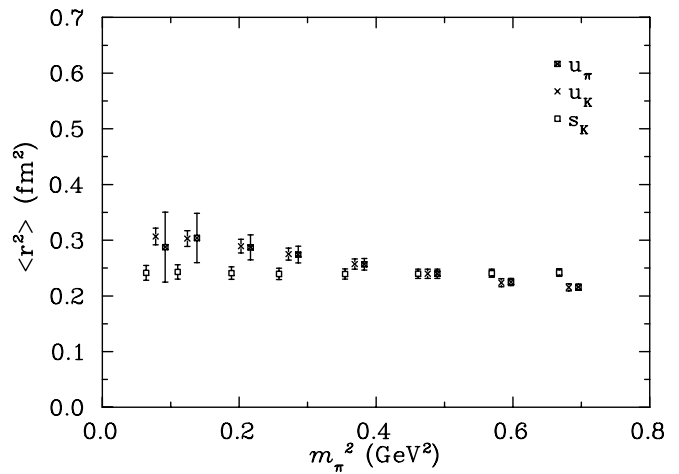}%
\caption{\label{fig:rsquaredpseudoscalarquarksectors}
The quark sector contributions to the mean squared charge radius of the pseudoscalar mesons.
The symbols are offset horizontally for clarity.}
\end{figure}

\begin{figure}[ht]
\includegraphics[height=1.0\hsize,angle=90]{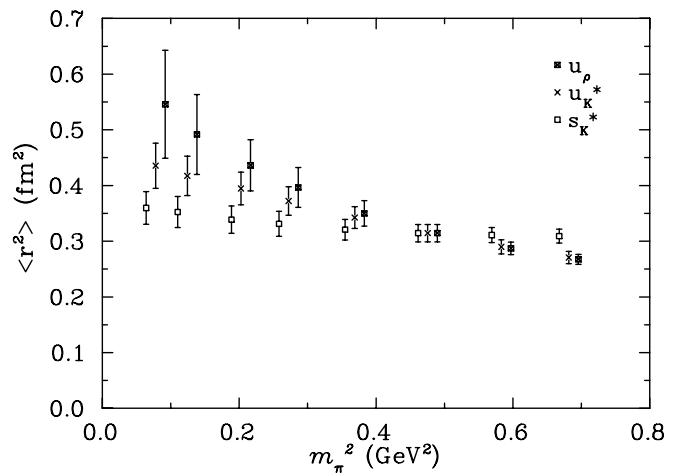}%
\caption{\label{fig:rsquaredvectorquarksectors}
As for Figure~\ref{fig:rsquaredpseudoscalarquarksectors} but for vector-mesons.
}
\end{figure}

By comparing the results for the up-quark contributions to the $\pi$ and $K$ ($\rho$ and $K^*$) charge radii, it is possible to gain insights into the effect that the presence of a heavier strange-quark has on the lighter up-quark in pseudoscalar (vector) mesons.
Figures~\ref{fig:rsquaredpseudoscalarquarksectors} and
\ref{fig:rsquaredvectorquarksectors} show the quark sector
contributions to the charge radii ($\langle r^2\rangle$) of the
pseudoscalar and vector mesons, respectively.  The quark sector
contributions to the charge radii for the pseudoscalar and vector
meson are recorded in Tables~\ref{tab:rsquared.pseudovector} and
\ref{tab:rsquared.vector}.
From Fig.~\ref{fig:rsquaredpseudoscalarquarksectors}, we find no
evidence of environmental sensitivity in the light-quark contribution
the pseudoscalar mesons.
However in the vector sector,
Fig.~\ref{fig:rsquaredvectorquarksectors}, we find a consistently
broader distribution of up-quark charge in the $\rho$ compared
to the up-quark in the $K^{*}$ at the smaller quark masses.
The broadening of the charge distribution in the $\rho$ is
consistent with the hyperfine repulsion discussed above.
The strange quark in the $K^{*}$ shows a particularly interesting 
environment sensitivity. While the strange quark mass is held 
fixed, the distribution broadens as the light-quark regime is 
approached. This is consistent with the prediction of enhanced 
hyperfine repulsion as one of the quarks becomes light.

\begin{table}[tbh]
\caption{\label{tab:rsquared.pseudovector} Mean-square charge radius ($\langle r^2 \rangle$) for quarks of unit charge in units 
of ${\rm fm}^2$. $m_\pi^2$ is given as a measure of the input quark mass.}
\begin{ruledtabular}
\begin{tabular}{llll}
$ m_\pi^2\ ({\rm GeV}^{2})$ & $u_\pi$ & $u_K$ & $s_K$ \\
\hline
$0.6956(26)$ &  $0.216(5)$   &  $0.215(7)$    &  $0.242(7)$ \\
$0.5970(26)$ &  $0.225(6)$   &  $0.224(7)$    &  $0.241(7)$ \\
$0.4895(24)$ &  $0.240(8)$   &  $0.240(8)$    &  $0.240(8)$ \\
$0.3828(23)$ &  $0.256(10)$  &  $0.257(9)$    &  $0.239(9)$ \\
$0.2862(25)$ &  $0.274(14)$  &  $0.275(11)$   &  $0.239(10)$ \\
$0.2166(27)$ &  $0.287(22)$  &  $0.289(12)$   &  $0.241(11)$ \\ 
$0.1382(48)$ &  $0.304(44)$  &  $0.303(14)$   &  $0.243(13)$ \\
$0.0920(49)$ &  $0.287(63)$  &  $0.306(15)$   &  $0.241(13)$ 
\end{tabular}
\end{ruledtabular}
\end{table}

\begin{table}[htb]
\caption{\label{tab:rsquared.vector} Mean-square charge radius ($\langle r^2 \rangle$) for quarks of unit charge in units 
of ${\rm fm}^2$.}
\begin{ruledtabular}
\begin{tabular}{llll}
$ m_\pi^2\ ({\rm GeV}^{2})$ & $u_\rho$ & $u_{K^*}$ & $s_{K^*}$ \\
\hline
$0.6956(26)$ &  $0.268(9) $  &  $0.271(11)$  &  $0.309(12)$ \\
$0.5970(26)$ &  $0.287(11)$  &  $0.290(13)$  &  $0.311(14)$ \\
$0.4895(24)$ &  $0.315(16)$  &  $0.315(16)$  &  $0.315(16)$ \\
$0.3828(23)$ &  $0.350(23)$  &  $0.342(20)$  &  $0.321(19)$ \\
$0.2862(25)$ &  $0.397(36)$  &  $0.372(26)$  &  $0.331(23)$ \\
$0.2166(27)$ &  $0.436(46)$  &  $0.395(30)$  &  $0.339(25)$ \\ 
$0.1382(48)$ &  $0.492(72)$  &  $0.417(35)$  &  $0.353(28)$ \\
$0.0920(49)$ &  $0.546(97)$  &  $0.436(41)$  &  $0.360(29)$ \\
\end{tabular}
\end{ruledtabular}
\end{table}

\begin{figure}[ht]
\includegraphics[height=1.0\hsize,angle=90]{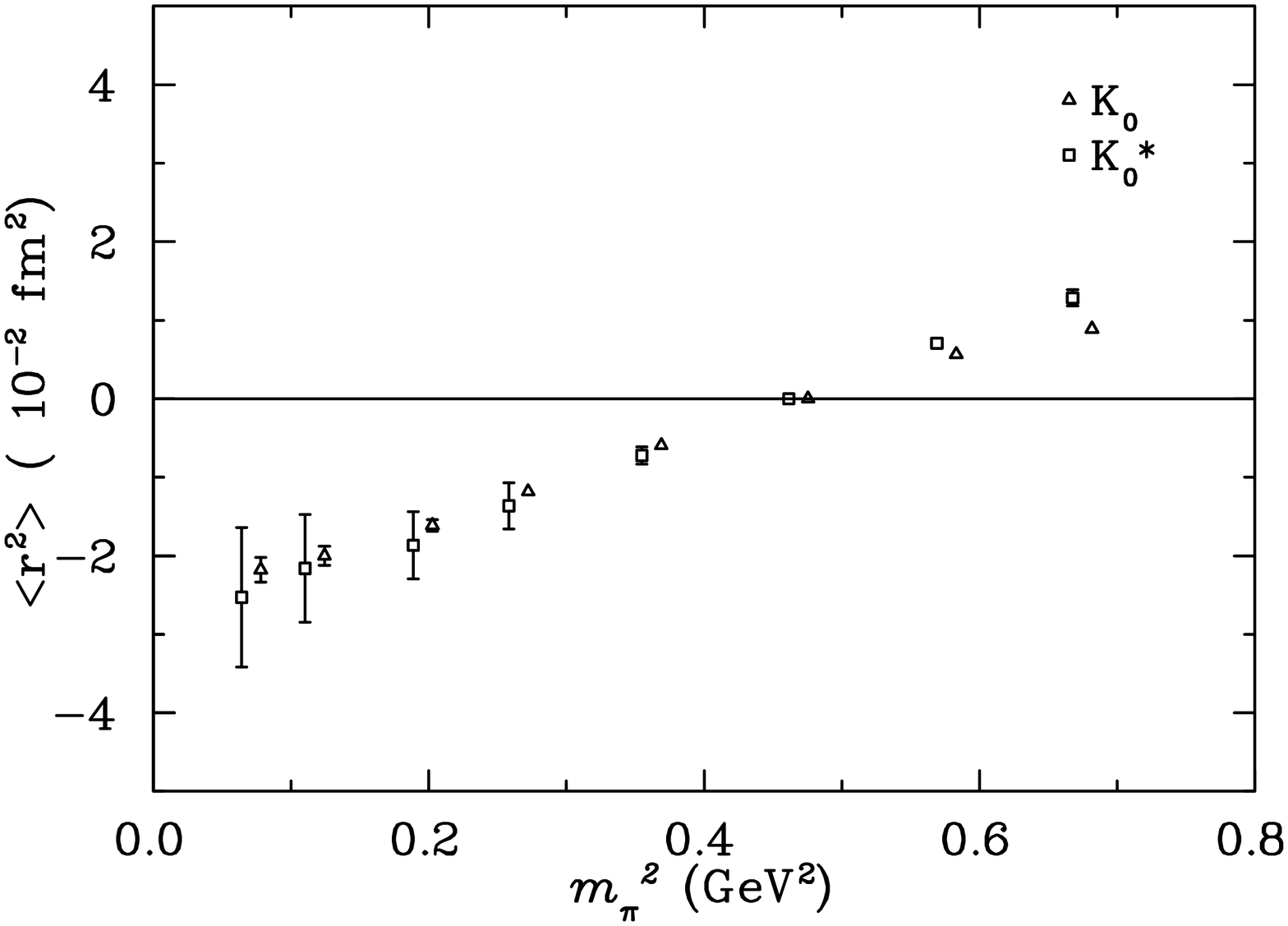}%
\caption{\label{fig:rsquareduncharged}
The mean squared charge radii for the neutral $K^{0}$ and $K^{0 *}$.
}
\end{figure}

The strange neutral pseudoscalar and vector meson mean squared charge
radii obtained from the weighted sum of the quark sector radii are
displayed in Fig.~\ref{fig:rsquareduncharged}.
For the neutral strange mesons, we see a negative value for $\langle r^2
\rangle$, indicating that the negatively charged $d$-quark is lying
further from the centre of mass on average than the $\bar{s}$. 
We should expect just such a behaviour for two reasons, both stemming from the fact that the $\bar{s}$ quark is considerably heavier than the $d$: the centre of mass must lie closer to the $\bar{s}$, and the $d$-quark will also
have a larger Compton wavelength.
Of course with exact isospin symmetry in our simulations, the non-strange charge neutral mesons have a zero electric charge radius.

\begin{figure}[t]
\includegraphics[height=1.0\hsize,angle=90]{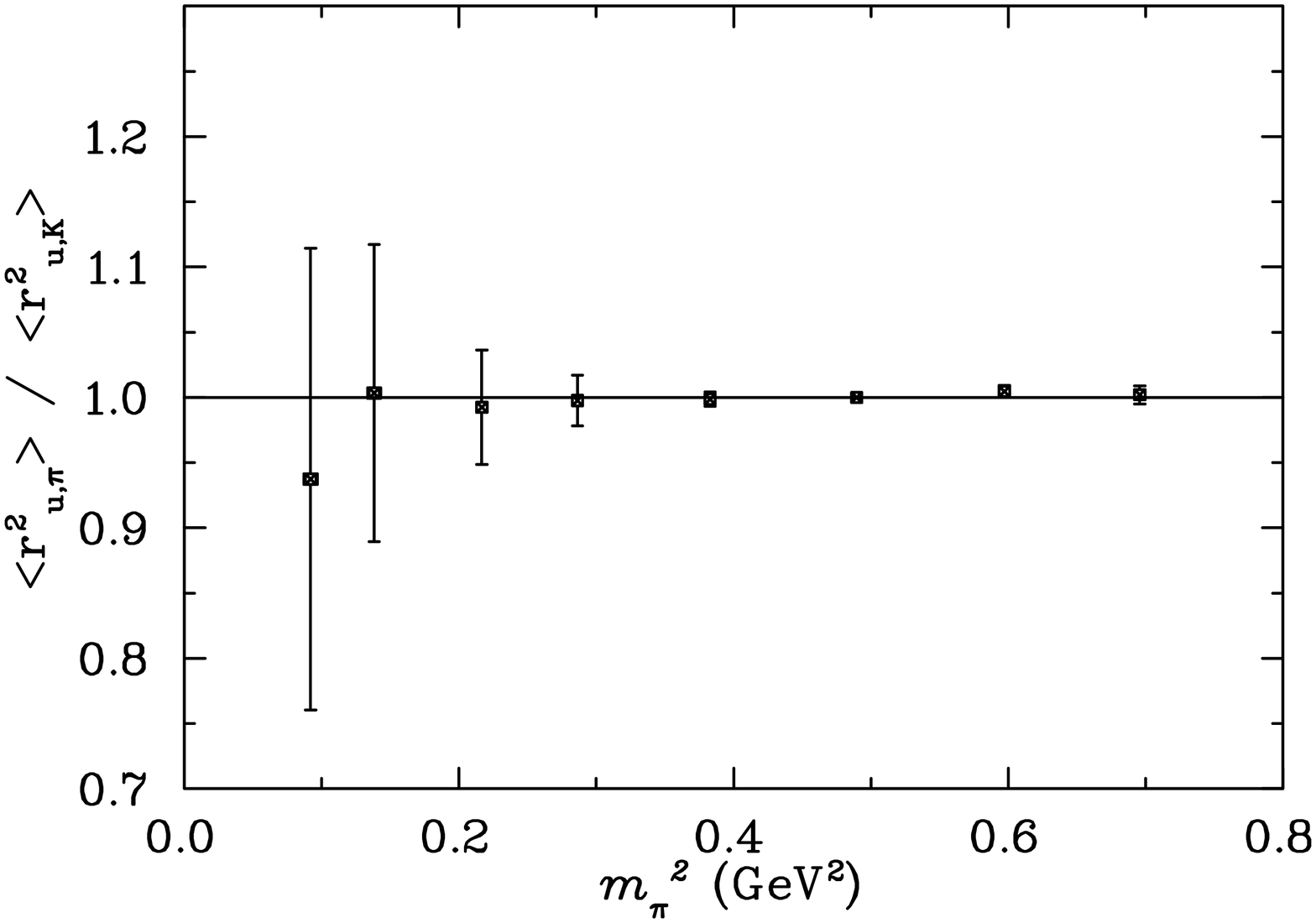}%
\caption{\label{fig:rsquaredenvironmentdependencepseudo}
The ratio of the light quark contributions to the $\pi$ and $K$ 
mean squared charge radius.
}
\end{figure}

\begin{figure}[t]
\includegraphics[height=1.0\hsize,angle=90]{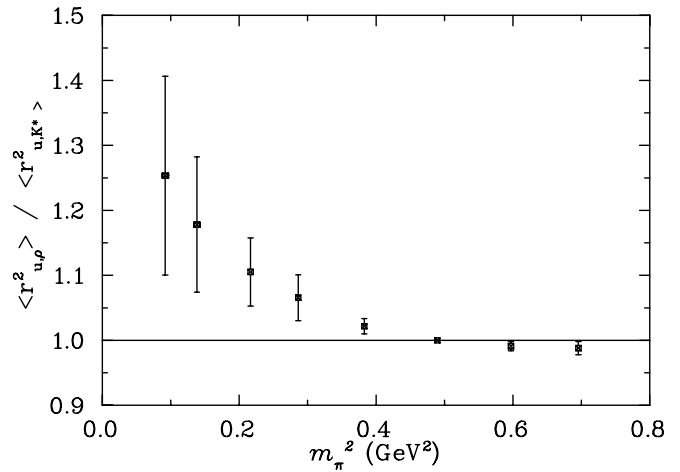}%
\caption{\label{fig:rsquaredenvironmentdependencevec}
As in Fig.~\ref{fig:rsquaredenvironmentdependencepseudo} but for the vector-mesons.
}
\end{figure}

To measure the environmental sensitivity of the light-quark sector
more precisely, in Figs.~\ref{fig:rsquaredenvironmentdependencepseudo}
and \ref{fig:rsquaredenvironmentdependencevec} we show a fit to the
ratio of the light-quark contributions to the pseudoscalar and
vector-mesons charge radii respectively.
%
%
The difference is striking: for the pseudoscalar case we see no
environment-dependence at all, whereas in the vector case we see that
the presence of a strange quark acts to heavily suppress the light
charge distribution.
This is the effect one predicts from a quark model, where the large
mass of the $s$ would act to suppress the hyperfine repulsion between
the quark and anti-quark. It is also qualitatively consistent with
effective field theory where the couplings of the light mesons are
suppressed by the presence of the strange quark.

\subsubsection{Magnetic moments}
\begin{figure}
\begin{center}
\includegraphics[height=1.0\hsize,angle=90]{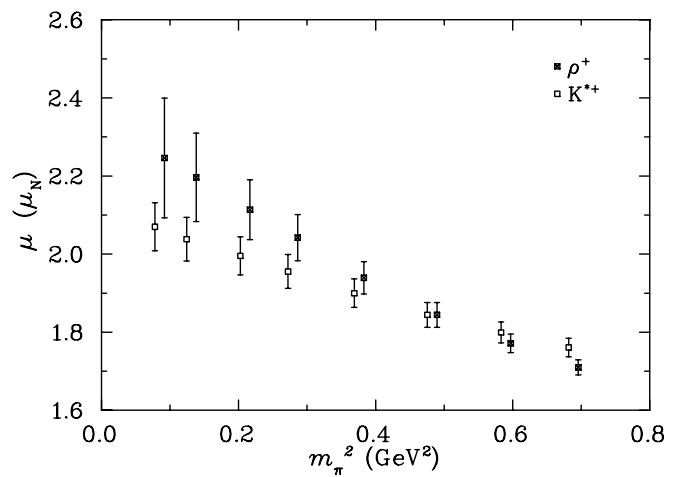}
\caption{\label{fig:magmomhadrons}
Charged vector meson magnetic moments.
}
\end{center}
\end{figure}

In Fig.~\ref{fig:magmomhadrons} we present our results for the magnetic moments of the vector mesons.
At the $SU(3)_{{\rm flavour}}$ limit, where we take the light quark
flavours to have the same mass as the strange quark, quark model arguments
suggest the magnetic moment for a $\rho^{+}$ should be -3 times the strange
magnetic moment of the $\Lambda$ (assuming no environmental dependence). 
According to the particle data group~\cite{Eidelman:2004wy}, the magnetic moment of the $\Lambda$ is $-0.613\, \mu_N$.
Therefore we would naively expect a value of $1.84\,\mu_N$ for the magnetic moment of the $\rho^+$, which is consistent with our findings.

\begin{figure}
\begin{center}
\includegraphics[height=1.0\hsize,angle=90]{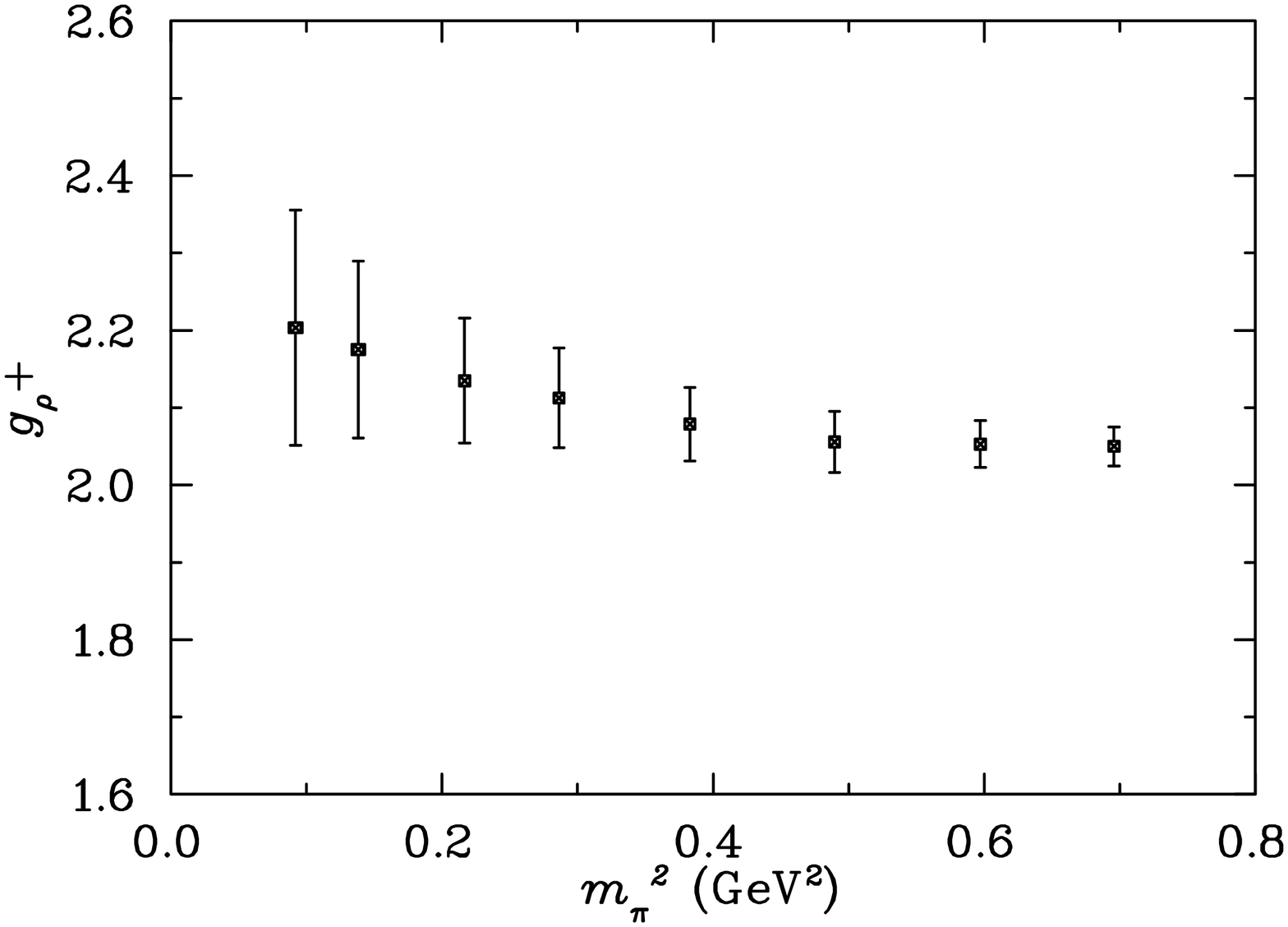}
\caption{\label{fig:gfactor.rhoplus}
The $g$-factor of the $\rho$ meson.
}
\end{center}
\end{figure}

In an earlier study, Anderson et al.~\cite{Andersen:1996qb} argued that the magnetic moment of the $\rho$-meson in natural magnetons (otherwise called the g-factor) should be approximately 2 at large quark masses.
Converting our result to natural magnetons, we observe in Fig.~\ref{fig:gfactor.rhoplus} that our calculation of the $\rho$-meson g-factor $(g_{\rho})$ is fairly consistent with this.
At light quark masses, however, we do see some evidence of chiral
curvature, which would indicate that the linear chiral
extrapolations of that paper should be considered with caution.

\begin{figure}
\begin{center}
\includegraphics[height=1.0\hsize,angle=90]{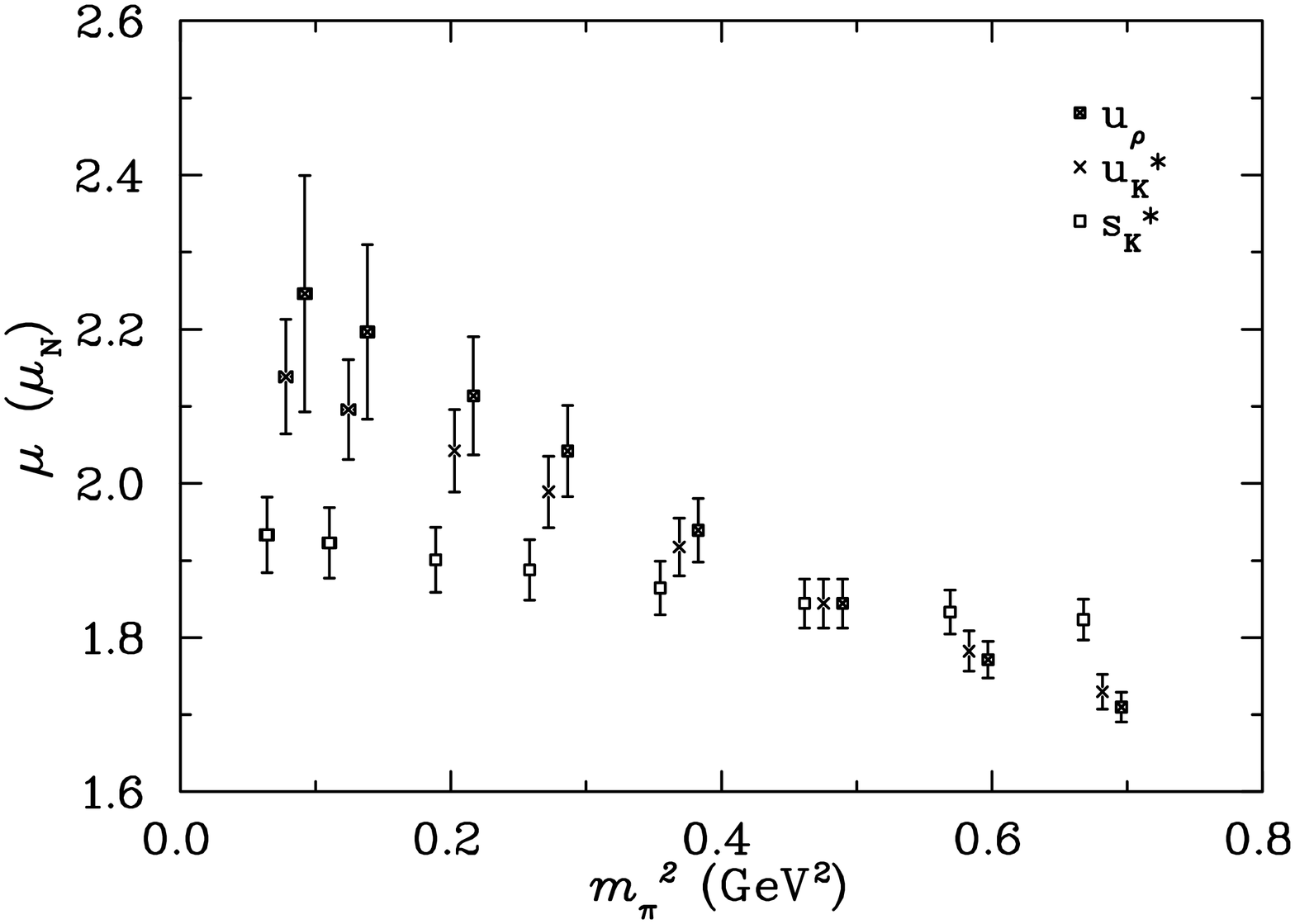}%
\caption{\label{fig:magmomquarksectors}
Quark-sector contributions to the vector meson magnetic moments.
}
\end{center}
\end{figure}

In Fig.~\ref{fig:magmomquarksectors} we present the quark sector contributions to the vector meson magnetic moments, the 
data is recorded in Table.~\ref{tab:magmom.vector}.
\begin{table}[htb]
\caption{\label{tab:magmom.vector} Magnetic moment for quarks of unit charge inside a vector meson in units 
of nuclear magnetons $\mu_{N}$.  }
  \begin{ruledtabular}
  \begin{tabular}{llll}
  $ m_\pi^2\ ({\rm GeV}^{2})$ & $u_\rho$ & $u_{K^*}$ & $s_{K^*}$ \\
  \hline
  $0.6956(26)$ &  $1.71(2)$  &  $1.73(2)$  &  $1.82(3)$ \\
  $0.5970(26)$ &  $1.77(2)$  &  $1.78(3)$  &  $1.83(3)$ \\
  $0.4895(24)$ &  $1.84(3)$  &  $1.84(3)$  &  $1.84(3)$ \\
  $0.3828(23)$ &  $1.94(4)$  &  $1.92(4)$  &  $1.86(3)$ \\
  $0.2862(25)$ &  $2.04(6)$  &  $1.99(5)$  &  $1.88(4)$ \\
  $0.2166(27)$ &  $2.11(8)$  &  $2.04(5)$  &  $1.90(4)$ \\ 
  $0.1382(48)$ &  $2.20(11)$ &  $2.10(6)$  &  $1.92(5)$ \\
  $0.0920(49)$ &  $2.25(15)$ &  $2.14(7)$  &  $1.93(5)$ \\
  \end{tabular}
  \end{ruledtabular}
\end{table}

\begin{figure}
\begin{center}
\includegraphics[height=1.0\hsize,angle=90]{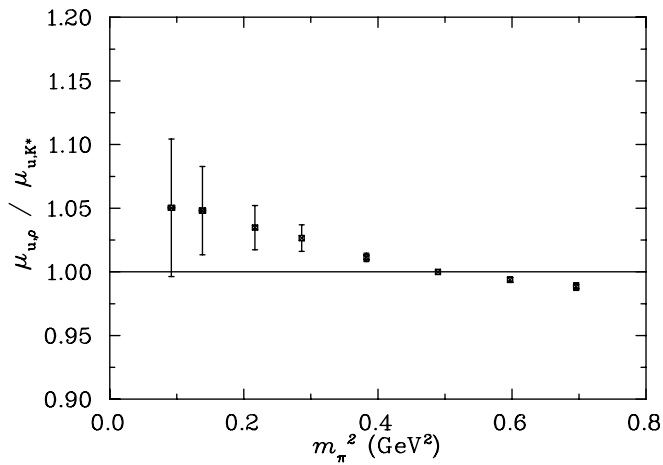}
\caption{\label{fig:magmomdependence}
The ratio of the light-quark contributions to the magnetic moment of the 
$\rho$ and $K^{*}$.
}
\end{center}
\end{figure}

Here we observe a similar scenario to that observed earlier in the
charge radius discussion, namely that the $u$-quark contribution to
the $K^*$ is consistently larger than the contribution from the heavier $s$-quark.
We also find that the contribution of the $u$-quark to the magnetic moment of a vector meson is suppressed when it is an environment of a heavier $s$-quark compared to when it is in the presence of another light quark.
This is further supported when we consider the ratio of the contributions of a $u$-quark to the magnetic moments of the $\rho$ and $K^*$ mesons, displayed in Fig.~\ref{fig:magmomdependence}.
This ratio is clearly greater than 1 below the $SU(3)_{\rm flavour}$ limit and is increasing for decreasing $u$-quark mass.

The magnetic moment of the vector meson, like the RMS charge radius, shows
considerable environment dependence in the quark sector contributions.
The larger contribution of a $u$-quark in a $\rho$ relative to a $K^*$ is
consistent with what we have already
observed with the RMS charge radius, as follows: since $\langle r^2 \rangle$ is
larger for the $u$-quark in a 
$\rho$ meson than for the $u$-quark in a $K^*$, the effective mass is
reciprocally smaller for the $u$-quark
in a $\rho$. This smaller effective mass gives rise in turn to a larger magnetic
moment. Figure \ref{fig:magmomdependence}
shows this pattern.
\begin{figure}
\begin{center}
\includegraphics[height=1.0\hsize,angle=90]{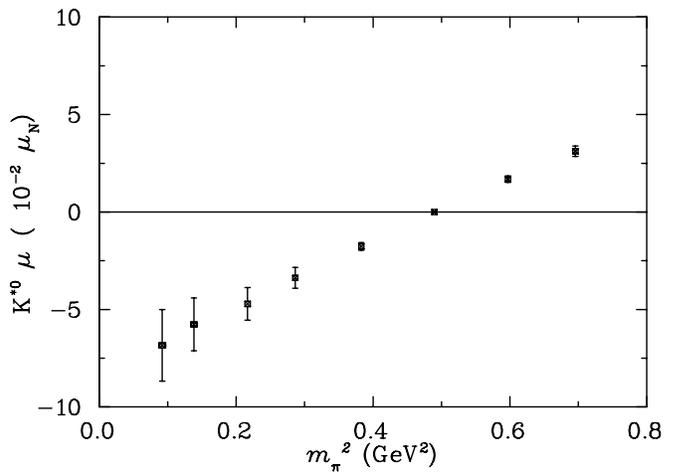}
\caption{\label{fig:magmomneutralkstar}
Neutral $K^*$-meson magnetic moment.
}
\end{center}
\end{figure}
Figure~\ref{fig:magmomneutralkstar} presents our results for the magnetic moment of the neutral $K^{*0}$ meson.
As the $d$-quark becomes lighter than the $\bar{s}$  we see the magnetic moment exhibiting a very linear negative slope. 
The magnitude of the magnetic moment is quite small, but clearly differentiable from zero everywhere except at the $SU(3)_{\rm flavour}$ limit where symmetry forces it to be exactly zero.
\subsubsection{Quadrupole form-factors}
\begin{figure}
\begin{center}
\includegraphics[height=1.0\hsize,angle=90]{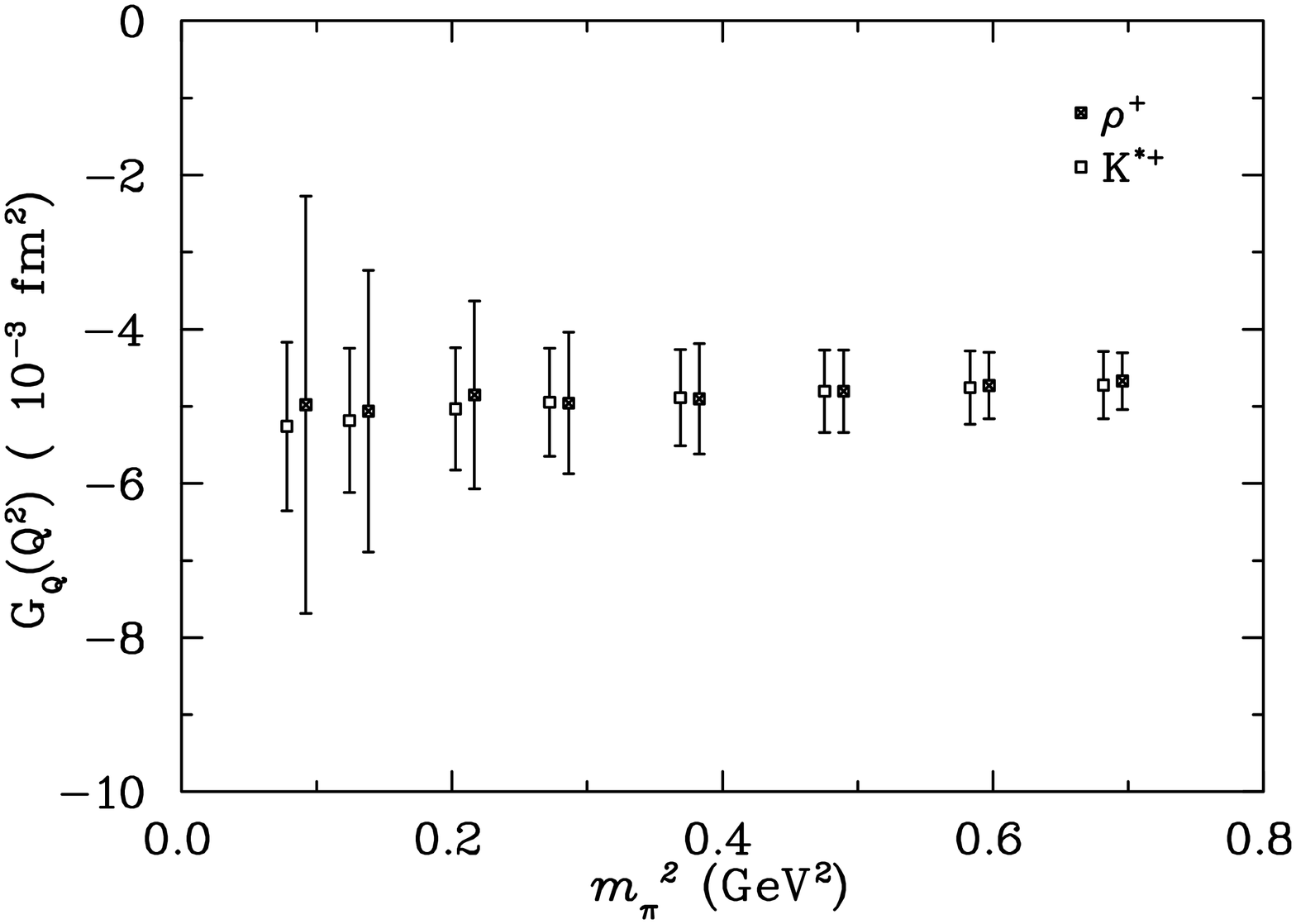}
\caption{\label{fig:quadffhadrons}
Vector meson quadrupole form factors for $\rho^+$ and $K^{*+}$.
}
\end{center}
\end{figure}

The quadrupole form-factors of the $\rho^+$ and $K^{*+}$ mesons are shown in Fig~\ref{fig:quadffhadrons}.
We find that the quadrupole form factor is less than zero, indicating that the spatial distribution of charge within the $\rho$ and $K^*$ mesons is oblate.
This is in accord with the findings of Alexandrou {\emph et al.}~\cite{Alexandrou:2002nn} who observed a negative quadrupole moment for spin $\pm 1$ $\rho$-meson states in a density-density analysis.
We note that in a simple quark model, a negative quadrupole form factor requires that the quarks possess an admixture of $s$- and $d$-wave functions.
\begin{figure}
\begin{center}
\includegraphics[height=1.0\hsize,angle=90]{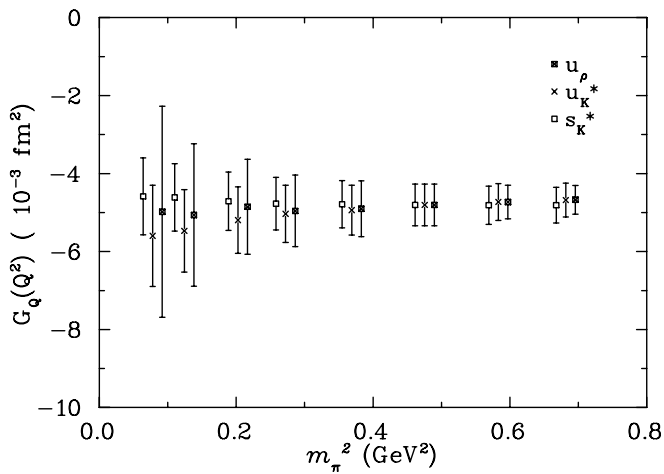}
\caption{\label{fig:quadffquarksectors}
Quark-sector contributions to the quadrupole form factors.
}
\end{center}
\end{figure}
\begin{table}[htb]
\caption{\label{tab:quadff-fm} The quadrupole form-factor (in ${\rm fm^2}$) for quarks of unit charge inside a vector meson.  }
\begin{ruledtabular}
\begin{tabular}{llll}
$ m_\pi^2\ ({\rm GeV})^{2}$ & $u_\rho$ & $u_{K^*}$ & $s_{K^*}$ \cr
\hline
$0.6956(26)$ &  $-0.0047(4)$  &  $-0.0047(4)$  &  $-0.0048(5)$ \cr
$0.5970(26)$ &  $-0.0047(4)$  &  $-0.0047(5)$  &  $-0.0048(5)$ \cr
$0.4895(24)$ &  $-0.0048(5)$  &  $-0.0048(5)$  &  $-0.0048(5)$ \cr
$0.3828(23)$ &  $-0.0049(7)$  &  $-0.0049(6)$  &  $-0.0048(6)$ \cr
$0.2862(25)$ &  $-0.0050(9)$  &  $-0.0050(7)$  &  $-0.0048(7)$ \cr
$0.2166(27)$ &  $-0.0049(12)$ &  $-0.0052(9)$  &  $-0.0047(7)$ \cr 
$0.1382(48)$ &  $-0.0051(19)$ &  $-0.0055(11)$ &  $-0.0046(9)$ \cr
$0.0920(49)$ &  $-0.0050(27)$ &  $-0.0056(13)$ &  $-0.0046(10)$ 
\end{tabular}
\end{ruledtabular}
\end{table}

The quark sector contributions to the quadrupole form-factor are shown
in Fig.~\ref{fig:quadffquarksectors}. The corresponding data is
contained in Table~\ref{tab:quadff-fm}. The flavour independence 
of the results is remarkable.
\begin{figure}
\begin{center}
\includegraphics[height=1.0\hsize,angle=90]{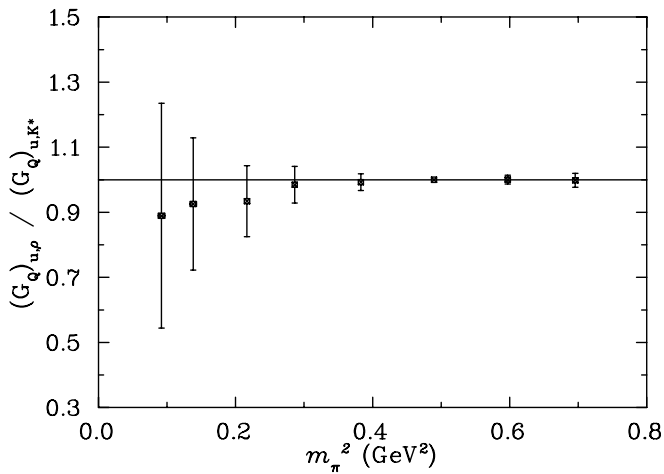}%
\caption{\label{fig:quadffdependence}
Environment-dependence for light-quark contribution to vector meson quadrupole
form-factor.
}
\end{center}
\end{figure}

We also find that the ratio of the light-quark contributions to the quadrupole form factor, 
shown in Fig.~\ref{fig:quadffdependence}, is consistent with one within our statistics.
\begin{figure}
\begin{center}
\includegraphics[height=1.0\hsize,angle=90]{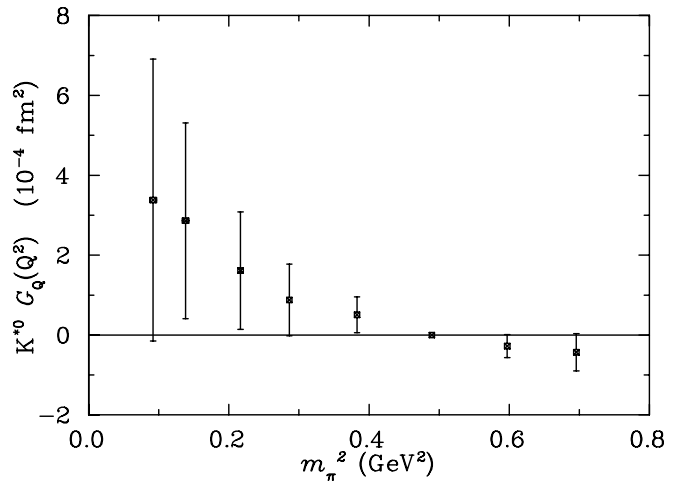}
\caption{\label{fig:quadffkstar0}
Quadrupole form-factor for neutral $K^*$ meson. 
}
\end{center}
\end{figure}
In Fig.~\ref{fig:quadffkstar0}, we show the quadrupole form factor of the charge neutral $K^{*0}$ meson.
We find that the quadrupole moment of the $K^{*0}$ is non-trivial but just outside the one 
standard deviation level. The chiral trend towards positive values reflecting the negative
charge of the larger $d$-quark contributions.

The lattice data for the quark sector contributions to the charge form factor is contained 
in Tables~\ref{tab:chargeff-ps} and ~\ref{tab:chargeff-vec} for the pseudoscalar and 
vector mesons respectively. The magnetic and quadrupole form factors of the vector 
mesons is contained in Tables~\ref{tab:magff} and \ref{tab:quadff} respectively.

\section{Conclusions}

We have established a formalism for determining the charge, magnetic and
quadrupole Sachs form factors of vector mesons in lattice QCD. For the
first time the electric, magnetic, and quadrupole form factors of the
light vector mesons have been calculated.  The electric form factor of
the pseudoscalar mesons have also been calculated.

With a large lattice volume and high statistics we have 
resolved a clear difference between the charge radii of the 
pseudoscalar and vector mesons. 
We argue that this is consistent with quark model predictions. 
Furthermore, we find significant environmental sensitivity of the light-quark contributions to the charge radii of the vector mesons.  

We also presented a calculation of the magnetic moments of the 
vector mesons. 
We found that the magnetic moment of the $\rho^{+}$ was
consistent with the quark model predication of $1.84\ \mu_N$ at the 
$SU(3)_{\rm flavour}$ limit. 
We determine that there is also an environmental sensitivity in the magnitude of the light-quark contributions to the charged vector meson magnetic moments. 
We argue that this is consistent with  the environmental sensitivity in the 
light-quark contributions to the charge vector meson charge radii.

Finally, we have determined that the quadrupole form factor for a
charged vector meson is negative in quenched Lattice QCD. This is
consistent with previous calculations using density-density
analysis. We find that the ratio of quadrupole moment to mean square
charge radius is 1:30, so the deformation is small but statistically
significant.
  
\section{Appendix}

\begin{table}[tbh]
\caption{\label{tab:chargeff-ps} The quark sector contributions to the charge form factor of the pseudoscalar mesons. }
\begin{ruledtabular}
\begin{tabular}{llll}
$ m_\pi^2\ ({\rm GeV})^{2}$ & $u_\pi$ & $u_K$ & $s_K$ \cr
\hline
$0.6956(26)$ & 0.833(4)  & 0.835(4) & 0.818(4) \cr
$0.5970(26)$ & 0.828(4)  & 0.830(5) & 0.820(5) \cr
$0.4895(24)$ & 0.822(5)  & 0.822(5) & 0.822(5) \cr
$0.3828(23)$ & 0.815(6)  & 0.813(6) & 0.823(6) \cr
$0.2862(25)$ & 0.810(8)  & 0.804(6) & 0.825(6) \cr
$0.2166(27)$ & 0.809(12) & 0.798(7) & 0.826(7) \cr 
$0.1382(48)$ & 0.812(22) & 0.792(8) & 0.826(8) \cr
$0.0920(49)$ & 0.833(30) & 0.791(8) & 0.828(8)
\end{tabular}
\end{ruledtabular}
\end{table}

\begin{table}[tbh]
\caption{\label{tab:chargeff-vec} As in Fig.~\ref{tab:chargeff-ps} but for the vector mesons. }
\begin{ruledtabular}
\begin{tabular}{llll}
$ m_\pi^2\ ({\rm GeV})^{2}$ & $u_\rho$ & $u_{K^{*}}$ & $s_{K^{*}}$ \\
\hline
$0.6956(26)$ & 0.795(5)  & 0.794(7)  & 0.771(7)  \cr
$0.5970(26)$ & 0.784(7)  & 0.783(8)  & 0.771(8)  \cr
$0.4895(24)$ & 0.769(9)  & 0.769(9)  & 0.769(9)  \cr
$0.3828(23)$ & 0.750(12) & 0.754(11) & 0.766(10) \cr
$0.2862(25)$ & 0.727(18) & 0.738(13) & 0.760(12) \cr
$0.2166(27)$ & 0.708(22) & 0.727(15) & 0.756(13) \cr
$0.1382(48)$ & 0.683(32) & 0.716(17) & 0.749(15) \cr
$0.0920(49)$ & 0.660(40) & 0.707(19) & 0.745(15)
\end{tabular}
\end{ruledtabular}
\end{table}

\begin{table}[tbh]
\caption{\label{tab:magff} The quark sector contributions to the magnetic form factor. }
\begin{ruledtabular}
\begin{tabular}{llll}
$ m_\pi^2\ ({\rm GeV})^{2}$ & $u_\rho$ & $u_{K^{*}}$ & $s_{K^{*}}$ \\
\hline
$0.6956(26)$ & 1.360(16) & 1.373(19) & 1.406(21) \cr
$0.5970(26)$ & 1.389(19) & 1.395(21) & 1.413(23) \cr
$0.4895(24)$ & 1.418(25) & 1.418(25) & 1.418(25) \cr
$0.3828(23)$ & 1.455(29) & 1.445(28) & 1.428(27) \cr 
$0.2862(25)$ & 1.484(37) & 1.469(31) & 1.435(28) \cr
$0.2166(27)$ & 1.496(47) & 1.485(36) & 1.438(30) \cr
$0.1382(48)$ & 1.500(60) & 1.501(41) & 1.440(31) \cr
$0.0920(49)$ & 1.483(81) & 1.513(46) & 1.441(33)
\end{tabular}
\end{ruledtabular}
\end{table}

\begin{table}[htb]
\caption{\label{tab:quadff} The quark sector contributions to the quadrupole form factor. }
\begin{ruledtabular}
\begin{tabular}{llll}
$ m_\pi^2 ({\rm GeV}^{2})$ & $u_\rho$ & $u_{K^*}$ & $s_{K^*}$ \cr
\hline
$0.6956(26)$ &  -0.285(22)  & -0.286(26) & -0.294(28) \cr
$0.5970(26)$ &  -0.289(26)  & -0.289(29) & -0.294(30) \cr
$0.4895(24)$ &  -0.293(33)  & -0.293(33) & -0.293(33) \cr
$0.3828(23)$ &  -0.299(44)  & -0.301(39) & -0.292(37) \cr
$0.2862(25)$ &  -0.303(56)  & -0.307(45) & -0.291(41) \cr
$0.2166(27)$ &  -0.296(74)  & -0.317(52) & -0.288(46) \cr
$0.1382(48)$ &  -0.309(112) & -0.334(65) & -0.282(53) \cr
$0.0920(49)$ &  -0.304(165) & -0.342(79) & -0.280(60)
\end{tabular}
\end{ruledtabular}
\end{table}

\begin{acknowledgments}

We thank the Australian Partnership for Advanced Computing
(APAC) and the South Australian Partnership for Advanced
Computing (SAPAC) for generous grants of supercomputer time which have
enabled this project.  This work was supported by the Australian
Research Council.
J.Z. is supported by PPARC grant PP/D000238/1.

\end{acknowledgments}

\bibliography{foo}

\end{document}